\documentclass[
	a4paper, %
	10pt, %
	unnumberedsections, %
	twoside, %
]{LTJournalArticle}

\usepackage{svg}

\AtEveryBibitem{
    \clearname{editor}
    \clearlist{publisher}
    \clearlist{location}
    \clearfield{day}
    \clearfield{month}
    \clearfield{note}
    \clearfield{pages}
    \clearfield{series}
    \clearfield{doi}
    \clearfield{isbn}
}
\runninghead{} %

\footertext{\textit{RISC-V Summit Europe, Bologna, 8-12th June 2026}} %

\title{CVA6-RT: an Open-Source Time-Predictable RV64 Processor for Mixed-Criticality Systems} %

\author{%
	Enrico Zelioli\textsuperscript{1}, Christopher Reinwardt\textsuperscript{1}, Nils Wistoff\textsuperscript{1}, Robert Balas\textsuperscript{1}\\ Alessandro Ottaviano\textsuperscript{3}, Luca Benini\textsuperscript{1,2}, Angelo Garofalo\textsuperscript{1,2}
}
\date{\footnotesize\textsuperscript{\textbf{1}}Integrated Systems Laboratory, ETH Zurich\\ \textsuperscript{\textbf{2}}Department of Electrical, Electronic, and Information Engineering, University of Bologna\\ \textsuperscript{\textbf{3}}Tenstorrent}

\renewcommand{\maketitlehookd}{%
	\begin{abstract}
        \noindent This work presents CVA6-RT, a real-time micro-architectural extension of the CVA6 core to bound worst-case latency and reduce task's timing execution variability. CVA6-RT implements the \emph{rv64gch} ISA and features advanced support for real-time execution, including TLB partitioning and locking for predictable address translation, a dynamically reconfigurable scratchpad mode in the L1 caches for deterministic memory access, and low-latency interrupt handling via an enhanced interrupt controller combined with hardware-assisted context stacking. With real-time features enabled, CVA6-RT achieves an interrupt latency of 12 cycles, comparable to that of simpler Arm Cortex-M microcontrollers, and 10$\times$ lower than the baseline CVA6 core.
	\end{abstract}
}

\begin{document}

\maketitle %

\renewcommand{\baselinestretch}{0.98}

\section{Introduction}

Mixed-criticality (MC) systems must execute workloads with different time-criticality levels, from performance-oriented to control tasks with bounded worst-case latency. The architectural trend in automotive and robotics toward centralizing more processing within a single SoC (e.g., zonal controllers) requires processors to simultaneously run operating systems, sensor-processing, and control tasks, making time-predictable execution challenging, as resource contention among tasks generates interference and latency variability.

Resource partitioning is essential to mitigate this problem. SW-only approaches such as cache and TLB coloring are often impractical due to high latency overhead~\cite{cachesurvey}. Partitioning the SoC into criticality domains (e.g., 32-b/64-b control/application processors) increases costs and reduces flexibility, possibly underutilizing one domain depending on the application's criticality profile. 
To overcome these limitations, a novel direction is to natively integrate real-time hardware support in 64-b processor architecture. A notable example is the Cortex-R class of ARM-based processors, integrated in several automotive zonal controllers, such as STMicroelectronics' Stellar SoCs~\cite{st_stellar}. 

However, the RISC-V ecosystem currently lacks an application-class processor with comparable real-time capabilities. We fill this gap by proposing CVA6-RT, which extends the micro-architecture of CVA6 with the following hardware-based time-predictability mechanisms:

\begin{itemize} 
\item Software-configurable, hardware-based TLB resource partitioning for predictable address translation latency;
\item Runtime-configurable L1 instruction and data cache resources in scratchpad mode for deterministic memory access latency; 
\item Deterministic low-latency interrupt handling via an enhanced RISC-V CLIC and hardware-assisted register stacking for fast context switch. \end{itemize}

\vspace{-0.5em}

Using interrupt latency as a representative use case, we show that CVA6-RT achieves 12 cycle interrupt latency, comparable to simpler ARM Cortex-M processors~\cite{cortexm}, and 10$\times$ lower than the baseline CVA6.

\begin{figure}[t]
    \centering
    \includegraphics[width=\columnwidth]{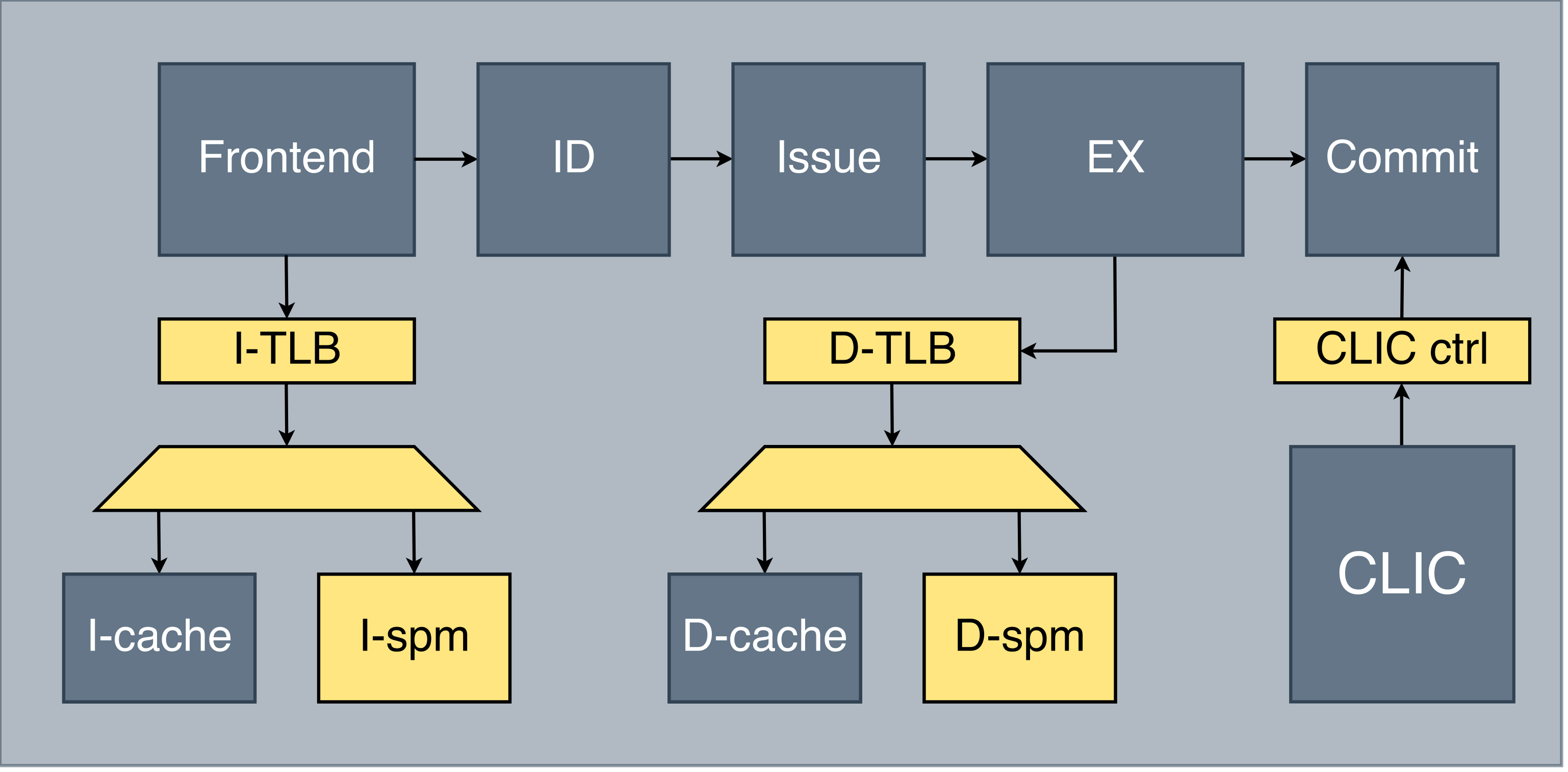}
    \caption{CVA6-RT block diagram with enhanced modules highlighted.}
    \label{fig:block_diagram}
    \vspace{-1.5em}
\end{figure}

\vspace{-1em}

\section{CVA6-RT Architecture}

CVA6~\cite{CVA6} is an open-source 64-bit RISC-V application-class core that supports virtual memory, private L1 instruction and data caches, and full operating system stacks such as Linux, widely used as a reference open-source application processor in both research and industry. Figure \ref{fig:block_diagram} shows an overview of CVA6-RT's micro-architecture described in this section, highlighting the changes with respect to vanilla CVA6.

\subsubsection{Predictable memory access latency}
We enable predictable memory access in CVA6 by introducing a hybrid cache/scratchpad (SPM) mode in the L1 instruction and data caches. Each cache way can be dynamically configured either as a conventional cache way or as software-managed scratchpad memory. Ways assigned to SPM are removed from the cache replacement logic, and their tags and valid bits are cleared to prevent unintended cache hits. Address decoding logic in the cache controllers maps designated physical address regions to the SPM, which is integrated into the SoC address space. The SPM region is organized as contiguous cache ways to simplify hardware mapping.
This design provides core-local memory with constant access latency, independent of cache replacement effects, allowing the operating system to guarantee bounded access latency to time-critical code and data, such as interrupt handlers and process stacks.

\begin{figure}[t]
    \centering
    \includegraphics[width=\columnwidth]{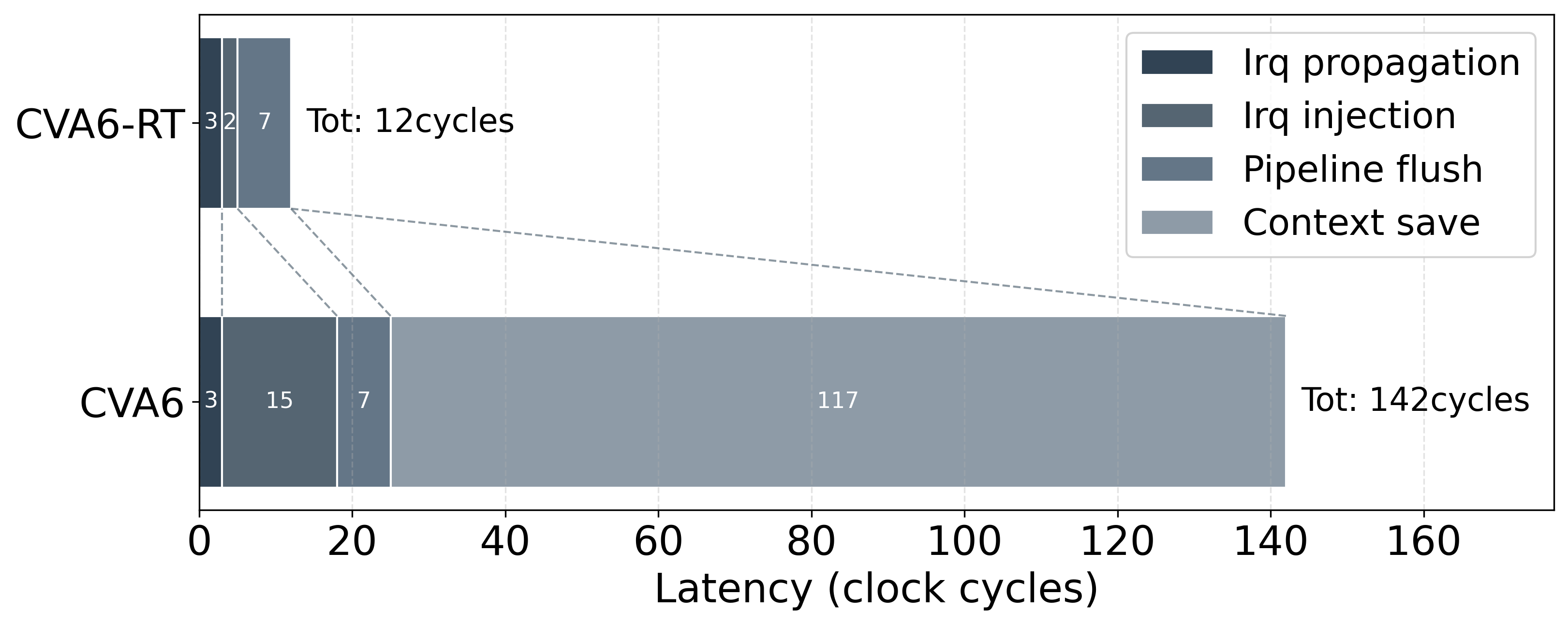}
    \caption{Average interrupt latency breakdown.}
    \label{fig:latency}
    \vspace{-2em}
\end{figure}

\subsubsection{Predictable address translation}

CVA6-RT extends the original pseudo-LRU-based TLB replacement logic to support partitioning of entries into configurable, non-overlapping sets. The pseudo-LRU tree is augmented with constraints that restrict which subsets of entries may be used for replacement. Privileged software selects active partitions via a bitmap, thereby controlling how many and which TLB entries a task can use. If tasks are assigned disjoint partitions, their TLB replacements are isolated, preventing eviction-based interference on a shared core. In addition, CVA6-RT provides a mechanism to statically lock a configurable number of TLB entries.
Once configured, these entries are removed from the replacement mechanism and cannot be evicted. This guarantees constant-latency address translation for selected memory regions, enabling predictable virtual memory access for time-critical tasks.

\subsubsection{Fast interrupt handling}

CVA6-RT reduces interrupt latency and jitter by integrating an enhanced version of the RISC-V Core-Local Interrupt Controller (CLIC), and extending the core's trap handling logic to support hardware-managed context save. The CLIC provides fine-grained interrupt prioritization and preemption, enabling low-latency interrupt vectoring, nesting, and tail-chaining. Moreover, the CLIC has been extended to support virtualization, enabling low, predictable interrupt latency for time-critical tasks also in virtualized systems.
To minimize software overhead during trap entry, CVA6-RT adds hardware-assisted context save support. Upon interrupt entry, the core can automatically spill a configurable subset of registers to a predefined memory location. 
This mechanism significantly shortens the interrupt latency and reduces variability introduced by software-managed register stacking. Combined with CLIC-based preemption and vectoring, these extensions enable bounded and low-latency interrupt handling suitable for real-time workloads in mixed-criticality systems.

\begin{table}
       \caption{Interrupt latency variability.}
       \centering
       \begin{tabular}{l r r r r}
               \toprule
               \multicolumn{1}{c}{Contribution} & 
               \multicolumn{2}{c}{CVA6} &
               \multicolumn{2}{c}{CVA6-RT}
               \\
               \cmidrule(r){2-3}
               \cmidrule(r){4-5}
               & Min & Max & Min & Max \\
               \midrule
               Irq propagation & 3 & 3 & 3 & 3 \\
               Irq injection & 6 & 30 & 2 & 3 \\
               Pipeline flush & 7 & 7 & 7 & 7 \\
               Context save & 104 & 130 & 0 & 0 \\
               \textbf{Total} & 120 & 170 & 12 & 13 \\
               \bottomrule
       \end{tabular}
       \vspace{-1em}
       \label{tab:latency}
\end{table}

\section{Evaluation and Conclusion}

As a case study, we evaluate the real-time suitability of CVA6-RT using interrupt latency. As shown in Fig.~\ref{fig:latency}, CVA6-RT achieves an average interrupt latency of 12 cycles, representing a 10$\times$ reduction compared to the 140-cycle latency of the baseline CVA6. The largest contribution comes from hardware-assisted register stacking, which removes the need for software-managed context saving and eliminates the associated variability. Interrupt synchronization latency is further reduced by relocating the interrupt detection logic from the instruction decode stage to the commit stage. In addition, trap entry overhead is reduced to approximately pipeline flush latency by enabling direct interrupt injection to lower privilege modes, including support for virtualized environments. As summarized in table \ref{tab:latency}, the interrupt latency variability of CVA6-RT is significantly reduced in all its components. Combined with predictable address translation and deterministic memory access, these improvements provide bounded and analyzable interrupt response times, making CVA6-RT suitable for soft real-time workloads in mixed-criticality systems.

\vspace{-1em}

\renewcommand{\baselinestretch}{0.9}

\printbibliography %

\end{document}